\documentclass{iopart}
\usepackage[dvips]{graphicx}
\begin{document}

\title[D-dimensional developed MHD turbulence: Double expansion model]
      {D-dimensional developed MHD turbulence: \\ Double expansion model}

\author{M Jurcisin \footnote{Present address: JINR Dubna, Moscow Region, Russia}
       and M Stehlik}

\address{Institute of Experimental Physics SAS, Ko\v{s}ice, Slovakia}
\eads{\mailto{jurcisin@thsun1.jinr.ru}, \mailto{stehlik@saske.sk}}

\begin{abstract}
Developed magnetohydrodynamic turbulence near two dimensions $d$ up
to three dimensions has been investigated by means of
renormalization group approach and double expansion regularization.
A modification of standard minimal subtraction scheme has been used
to analyze the stability of the Kolmogorov scaling regime which is
governed by the renormalization group fixed point. The exact
analytical expressions have been obtained for the fixed  points.
The continuation of the universal value of the inverse Prandtl number
$u=1.562$ determined at $d=2$ up to $d=3$ restores the value of $u=1.393$
which is known in the kinetic fixed point from usual $\epsilon$-expansion.
The magnetic stable fixed point has been calculated and its
stability region has been also examined.
This point losses stability: (1) below critical value of dimension
$d_c=2.36$ (independently on the $a$-parameter of a magnetic
forcing) and, (2) below the value of $a_c=0.146$ (independently on
the dimension).
\end{abstract}
\pacs{ 47.27.ef, 52.65.Kj, 11.10.Hi}  %47.27.$-$i,
\submitto{\JPA}
%\maketitle

\section{Introduction}

The renormalization group (RG) methods have been widely used to
the analysis of fully developed hydrodynamic (HD) turbulence
beginning from pioneering papers \cite{FoNeSt77,DomMar79} based on
\cite{MaSiRo73,BaJaWa76}. It gives possibility to reply upon some
principal questions, e.g., on the fundamental description of the
infrared (IR) scale invariance, as well as it is useful for
calculation of many quantities, e.g., critical dimensions of the
fields and their gradients, viscosity, etc. (see, e.g.
\cite{AdAnVa96-99,Vasiliev98-04,McComb90-95,Davidson04}).

Then many authors begin to use Wilson's scheme or some adequate
generalized renormalization scheme to study of HD turbulence
\cite{YakOrs86-87} as well as of magnetohydrodynamic (MHD)
turbulence \cite{FoSuPo82,CamTas92}. This time Vasiliev's team
have used functional formulation of the field-theoretic RG
\cite{Vasiliev76,Zinn89} to legalize of the Kolmogorov scaling
regime of HD turbulence \cite{AdVaPi83,AdVaHn84}. They consider
(as used in present paper) the functional quantum field RG
approach \cite{AdAnVa96-99} rather then Wilson's RG technique
\cite{WilKog74}. It assigns a field action to the stochastic
problem and makes possible to use elegant and very well developed
RG procedure in quantum field theory to investigate infrared
asymptotic regimes of a stochastic system. Then this
RG method has been applied in MHD turbulence
\cite{AdVaHn85,AdVaHn87}. Note here that this functional RG method
allow a straightforward extension of the pertubative calculation
to an higher order loops without a principal difficulty (see
\cite{AdAnKoVa03,AdAnBa01}, for example).

Considerable effort had been devoted to application of adequate field-theoretical
methods in the MHD turbulence (see recent review of Verma \cite{Verma04}, for example).
Authors in \cite{FoSuPo82,CamTas92} have used the 'classical' Yakhot-Orszag scheme
\cite{YakOrs86-87}.
In last years Verma \cite{Verma01a,Verma01b} performed detailed RG calculation of
MDH turbulence using McComb's alternative field-theoretic RG procedure
\cite{McComb90-95} and they reached notable progress in calculation of some
renormalized parameters of MHD turbulence.
Here we will not present full discussion of all methods used in the full developed
turbulence theory such as calculation of Alfven ratio, magnetic resistivity
\cite{Verma01b} or a problem of magnetic dynamo in helicit MHD \cite{AdVaHn87}
because it goes out of the frame of present paper (but see some remarks and
discussion in section 4 and section 5).
%Here we restrict ourselves to the remark that us used RG method fixes already
%only stable scaling regimes with a balanced kinetic and/or magnetic energy density
%in stationary turbulent regime due to consistent problem formulation.

Present paper deals with an investigation of the existence and range of stability
of the 'magnetic' scaling regime (i.e. the magnetic fixed point for zero
inverse Prandtl number, see below) in the non-helical $d$-dimensional MHD turbulence.
%%%==
The existence of two different anomalous scaling regimes in three dimensions,
which are known as kinetic and magnetic ones, was established in the pioneering
papers \cite{FoSuPo82,AdVaHn85}. These two points correspond to two
IR stable fixed points of the RG. On the other hand, it was also
supposed that in two dimensions the magnetic fixed point does not
exist as a result of nonexistence of the IR stable magnetic fixed
point. But the conclusions about  two dimensional fixed points
cannot be consider without doubts in these papers due to the
problems with renormalization in two dimensions which were not taken
into account \cite{HonNal96} (see also \cite{AdAnVa96-99}). In
\cite{Liang93,Kim99} two dimensional case was studied too but again
with shortcomings, therefore their results cannot be considered
completely conclusive. Within the our field theoretic RG approach the
problem is related to the existence of additional divergences which
arise in two dimensions.

The first correct treatment of the two dimensional case of the
stochastically forced MHD equations with the proper account of these
additional divergences was done in \cite{HnHoJu01}. It was
accomplished within a two-parameter expansion (double expansion) of
scaling exponents and scaling functions \cite{HonNal96} where,
besides the parameter which characterize the deviation of the
exponent of the powerlike correlation function of random forcing
from its critical value, the additional parameter of the deviation
of the spatial dimension was introduced. The using of this double
expansion method has allowed them to confirm the basic conclusions
of the previous works \cite{FoSuPo82,AdVaHn85}, namely,  the
nonexistence of the magnetic scaling regime near two dimensions.

The authors of the paper \cite{HnHoJu01} also tried to restore the
stability of the magnetic fixed point when moving from two
dimensions in direction of three dimensions. This possibility was
achieved by using of the special choice of finite renormalization
which allowed them to keep track of the effect of the additional
divergences near two dimensions. Technically, it was done by
introducing of another uniform UV cutoff in all propagators which
does not affect the large-scale properties of the model. This setup
is similar to that of Polchinski \cite{Polchinski}. As a result, the
borderline dimension between stable and unstable magnetic fixed
point was found and it leads to the possibility of the uniform
description of two and three dimensional cases of stochastic MHD.

Another possibility how to solve the problem of the additional
divergences in two dimensions together with the problem of
restoration of the stability of the corresponding fixed point when
going from a two dimensional system to a three dimensional one was
proposed in \cite{HnJoJuSt01}. They suggest to apply a modified
minimal substraction (MS) scheme in which the $d$-dependence of the
tensor structures of the UV divergent parts of the corresponding
diagrams are kept. It was successfully used in the fully developed
Navier-Stokes turbulence with weak uniaxial anisotropy to restore
the stability of the Kolmogorov scaling regime which is unstable in
two dimensions and stable in three dimensions.

In what follows, we shall apply the double expansion method together
with modified MS scheme introduced in \cite{HnJoJuSt01}  to the
stochastic MHD equations. Our aim is to investigate if it is
possible to describe correctly and uniformly the two dimensional and
the three dimensional systems and to compare our results with that
of \cite{HnHoJu01} where the different method was used (see above).
Thus, we carry out an analysis of the randomly forced MHD equations
with the proper account of the additional UV divergences which are
appeared in $d=2$. We apply a modified minimal subtraction scheme
based on the fact that the tensor structure of counter-terms is left
generally $d$-dependent in the calculations of divergent parts of
Green's functions. It will be shown that it allows us to investigate
behavior of the system under continual transition to $d=3$ beginning
from $d=2$. We have also confirmed the earlier conclusions made in
\cite{FoSuPo82,AdVaHn85,Verma01a}
that near two dimensions a scaling regime
driven by the velocity fluctuations may exist, but no magnetically
driven scaling regime can occur. We have also investigated the
long-range asymptotic behavior of the model in the double expansion
framework and found, in particular, that in this case thermal
fluctuations of the magnetic scaling regime may occur and that the
value of the borderline dimension is significantly lower
($d_c=2.36$) than in the $\epsilon$ expansion \cite{FoSuPo82}
($d_c=2.85$) and rather lower than in the 'modified' double
expansion introduced in \cite{HnHoJu01} ($d_c=2.46$)
but it is rather higher then value ($d_c=2.2$) calculated in the frame of
the McComb's renormalization \cite{Verma04}.
The discrepancy between the value of inverse Prandtl number $u$ which corresponds
to nontrivial stable fixed point of the RG in the three dimensions, which has been
obtained in the double expansion scheme in earlier paper \cite{HnHoSt94} and that
obtained by the usual $\epsilon$-expansion scheme \cite{FoSuPo82,AdVaHn85}
and also that obtained by Verma \cite{Verma01a,Verma01b} by McComb's procedure,
was one more reason of the present analysis.
Here we show
that the continuous transition from the universal value of the inverse Prandtl
number $u=1.562$ determined at $d=2$ restores the value of $u=1.393$ at $d=3$
which is known from usual $\epsilon$-expansion.

The paper is organized as follows: In section~\ref{sec2} the functional
field theoretic formulation of the model is present in detail. In
section~\ref{sec3} the renormalization of the model is discussed. In
section~\ref{sec4} detailed analysis of the possible scaling regimes
is done. In section~\ref{sec5} conclusions and discussion of the
results are given.

\section{Functional formulation of double expansion model}\label{sec2}

In the present paper we study the universal statistical features of
the model of stochastic MHD which is described by the system of
equations for the fluctuating velocity field of an incompressible
conducting fluid $\bi{v}(x)$, $x\equiv (\bi{x},t)$,
$\bi{\nabla}\bi{\cdot}\bi{v} = 0$ and the magnetic induction ${\bf
B}=(\rho \mu)^{1/2} \bi{b}(x)$ (where $\rho$ is density of the fluid
and $\mu$ is its permeability)  \cite{FoSuPo82,AdVaHn85,HnaSte91}:
\begin{eqnarray}
 \partial_t \bi{v} + (\bi{v}\bi{\cdot}\bi{\nabla})\bi{v} -
(\bi{b}\bi{\cdot}\bi{\nabla})\bi{b} - \nu_0 \nabla^2 \bi{v} =
\bi{f^v} \,,
\label{NSR} \\
 \partial_t \bi{b} + (\bi{v}\bi{\cdot}\bi{\nabla}) \bi{b} -
(\bi{b}\bi{\cdot}\bi{\nabla})\bi{v} -\nu_0 u_0\nabla^2\bi{b}
=\bi{f^b}\,, \label{magnet}
\end{eqnarray}
with the incompressibility conditions $\bf{\nabla}\bi{\cdot}\bi{f^v}
=0$ and $\bf{\nabla}\bi{\cdot}\bi{f^b} =0$ and the field $\bi{b}$ is
suppose to be solenoidal too, $\bi{\nabla}\bi{\cdot}\bi{b}=0$. The
statistics of $\bi{v}$, $\bi{b}$  is completely determined by both
the non-linear equations (\ref{NSR},\ref{magnet}) and the statistics
of the external inter-correlated large-scale random forces
$\bi{f^v}$, $\bi{f^b}$. The dissipation is controlled by the
parameter of kinematic viscosity $\nu_0$, and $u_0$ denotes inverse
Prandtl number (hereafter all parameters with a subscript $0$ denote
bare parameters of unrenormalized theory; see below). Note here that
the term $(\bi{b}\bi{\cdot}\bi{\nabla})\bi{b}$ expresses the
transverse part of Lorentzian force and it can be omitted in the
case of magnetic field treated as a passive admixture.

As usual \cite{FoSuPo82,AdVaHn85}, statistical properties of the
Gaussian forcing with zero mean values
($\langle\,\bi{f^v}\,\rangle=0, \,\, \langle\,\bi{f^b}\,\rangle=0$)
are determined by relations:
\begin{eqnarray}\fl\quad
\langle \,f_j^v(1) f_s^v(2)\,\rangle = \delta(\tau) u_0\,\nu_0^3\,
\int\frac{\rmd^d\bi{k}}{(2\pi)^d}\,{\rm P}_{js}(\bi{k})
\rme^{\rmi\bi{k.x}} \left[ g_{v10}\, k^{2- 2\delta-2\epsilon} +
g_{v20}\,k^2\,\right] \,,
\label{corelv} \\
\fl\quad \langle\, f_j^b(1) f_s^b(2)\,\rangle = \delta(\tau)
u_0^2\,\nu_0^3\, \int\frac{\rmd^d\bi{k}}{(2\pi)^d}\,{\rm
P}_{js}(\bi{k}) \rme^{\rmi\bi{k.x}} \left[ g_{b10}\, k^{2- 2\delta-2
a\epsilon} + g_{b20}\,k^2\,\right] \,, \label{corelb}
\end{eqnarray}
where the argument $1\equiv x_1$, $\tau=t_1 - t_2$,
$\bi{x}=\bi{x_1}-\bi{x_2}$, ${\rm P}_{js}(\bi{k})=\delta_{js}-k_j
k_s/k^2 $, the parameter $\epsilon$ determines the powerlike falloff
of the long-range forcing correlations, and the parameter
$\delta=(d-2)/2$ describe the deviation from spatial dimension
$d=2$. The free parameter $a$ controls the power form of magnetic
forcing. Note that parameters $\epsilon=2, a=1$ are the natural
"physical" values in our "massless" power-law energy injection. The
introduction of the local correlations (proportional to the new
couplings $g_{v20}$, and $g_{b20}$) which are described by the
analytic in $k^2$ terms in the correlation functions (\ref{corelv}),
and (\ref{corelb}) is related to the existence of additional
divergences of this structure (see below in the text) in the two
dimensional model which cannot be removed by corresponding nonlocal
terms \cite{HonNal96,Honkonen98,HnHoHoSe99}. At the same time, the
localness of the counterterms is the fundamental feature of a model
to be multiplicatively renormalizable \cite{Collins85,Zinn89}. For
example, it was not taken into account in the analysis of the model
in \cite{FoSuPo82,AdVaHn85}.

Using the well-known Martin-Siggia-Rose formalism
\cite{MaSiRo73,BaJaWa76}, one can transform the stochastic problem
(\ref{NSR})-(\ref{magnet}) with correlators (\ref{corelv}), and
(\ref{corelb}) into the field theoretic model of the doubled set of
fields $\Phi\equiv\{\bi{v},\bi{b},\bi{v^{\prime}},\bi{b^{\prime}}\}$
with the following action functional
\begin{eqnarray}\fl
\quad S  = {1\over2}\int\rmd x_1 \rmd x_2\,
 \Big\{ v^{\prime}_j(1)\,\langle\,f_j^v(1)f_s^v(2)\,\rangle_0\,v^{\prime}_s(2)
    +   b^{\prime}_j(1)\,\langle\,f_j^b(1)f_s^b(2)\,\rangle_0\,b^{\prime}_s(2) \Big\}+
\nonumber\\
 +  \int\rmd x\ \bi{v^{\prime}} \bi{\cdot} \left( -\partial_t\bi{v} +
     \nu_0\,\nabla^2\bi{v} -(\bi{v}\bi{\cdot}\bi{\nabla})\bi{v}
     -(\bi{b}\bi{\cdot}\bi{\nabla})\bi{b}  \right)
\nonumber\\
 +  \int\rmd x\ \bi{b^{\prime}}\bi{\cdot}\left(-\partial_t\bi{b} +u_0\,\nu_0\nabla^2\bi{b}+
  (\bi{b}\bi{\cdot}\bi{\nabla})\bi{v}
  -(\bi{v}\bi{\cdot}\bi{\nabla})\bi{b}\right)\,,
 \label{action}
 \end{eqnarray}
where $\bi{v^{\prime}}$, and $\bi{b^{\prime}}$ are independent of
$\bi{v}$, and $\bi{b}$ auxiliary incompressible fields, which we
have to introduce when transforming the stochastic problem into a
functional form.

The dimensional constants $g_{v10}, g_{b10}, g_{v20}$, and
$g_{b20}$, which control the amount of randomly injected energy
given by (\ref{corelv}), (\ref{corelb}), play the role of the
coupling constants in the perturbative expansion.
%Their universal values have been determined after the parameters
%$\epsilon,\delta$ have been chosen to give the desired power form of
%forcing and desired dimension.
For the convenience of further calculations the factors $\nu_0^3
u_0$ and $\nu_0^3 u_0^2$ including the "bare" (molecular) viscosity
$\nu_0$ and the "bare" (molecular or microscopic) magnetic inverse
Prandtl number $u_0$ have been extracted. As was  mentioned already
the bare (non-renormalized) quantities are denoted by subscript "0".

The most important measurable quantities in the study of a fully
developed turbulence and related problems are considered to be the
statistical objects represented by correlation and response
functions (Green functions) of the fields. Standardly, the
formulation through the action functional (\ref{action}) replaces
the statistical averages of random quantities in the stochastic
problem (\ref{NSR})-(\ref{corelb}) with equivalent functional
averages with weight $\exp S(\Phi)$. Generating functionals of total
Green functions G(A) and connected Green functions W(A) are then
defined by the functional integral
\begin{equation}
G(A)=e^{W(A)}=\int {\cal D}\Phi \,\, e^{S(\Phi) +
A\Phi},\label{green}
\end{equation}
where $A(x)=\{{\bf A^{v}}, {\bf A^{b}}, {\bf A^{v^\prime}},{\bf
A^{b^{\prime}}}\}$ represents a set of arbitrary sources for the set
of fields $\Phi$, ${\cal D}\Phi \equiv {\cal D}\theta{\cal
D}\theta^{\prime}{\cal D}{\bf v}{\cal D}{\bf v^{\prime}}$ denotes
the measure of functional integration, and linear form $A\Phi$ is
defined as
\begin{equation}
\hspace{-2cm} A \Phi= \int d\,x [{\bf A^{v}}(x)\cdot {\bi v}(x)+
{\bf A^{b}}(x)\cdot {\bi b}(x) + {\bf A^{v^{\prime}}}(x)\cdot {\bi
v^{\prime}}(x) + {\bf A^{b^{\prime}}}(x)\cdot {\bi
b^{\prime}}(x)].\label{form}
\end{equation}

The functional formulation gives the possibility of using the field
theoretic methods, including the RG technique to solve the problem.
By means of the RG approach it is possible to extract large-scale
asymptotic behavior of the correlation functions after an
appropriate renormalization procedure which is needed to remove UV
divergences. The functional formulation is advantageous also because
the Green functions of the  Fourier-decomposed stochastic MHD can be
calculated by means of Feynman diagrammatic technique.

Action (\ref{action}) is given in a form convenient for a
realization of the field theoretic perturbation analysis with the
standard Feynman diagrammatic technique. Free (bare) propagators
${\hat\Delta}$ can be found from the quadratic part of the action
(\ref{action}) written in the form $\,\,-(1/2)\Phi\hat{{\cal
K}}\Phi$ and by using the definition $\,\,\hat{{\cal
K}}{\hat\Delta}=\hat1$, where $\hat1$ denotes the diagonal matrix
whose diagonal elements are the transverse projectors (our fields
are solenoidal). One obtains
 \begin{eqnarray}
 {\hat\Delta}_{js} &=&
 \left(
 \begin{array}{cccc}
\Delta^{vv}_{js} &      0      & \Delta^{vv'}_{js} &      0        \\
0            & \Delta^{bb}_{js} &   0          & \Delta^{bb'}_{js} \\
\Delta^{v'v}_{js} &     0        &    0         &        0        \\
0             & \Delta^{b'b}_{js} &     0        &       0
 \end{array} \right)
 \end{eqnarray}
with the elements (wave-number-frequency representation)
\begin{eqnarray}
\Delta^{vv'}_{js}({\bf k},\omega)&=& \Delta^{v^{\prime}v}_{js}(-{\bf
k},-\omega) =\frac{{\rm P}_{js}({\bf k})}{-i\omega+\nu_0\, k^2}\,,
\nonumber\\
\Delta^{bb'}_{js}({\bf k},\omega)&=& \Delta^{b^{\prime}b}_{js}(-{\bf
k},-\omega) =\frac{{\rm P}_{js}({\bf k})}{-i\omega+u_0\,\nu_0\,
k^2}\,,
\nonumber\\
\Delta^{vv}_{js}({\bf k},\omega) &=&u_0\,\nu_0^3\,k^2\
\frac{g_{v10}\,k^{-2\delta-2\epsilon}+g_{v20}}
      {|-i\omega+\nu_0\, k^2|^2}\,{\rm P}_{js}({\bf k})\,,
\nonumber\\
\Delta^{bb}_{js}({\bf k},\omega)&=&u_0^2\,\nu_0^3\,k^2 \
\frac{g_{b10}\,k^{-2\,a\delta-2\epsilon}+ g_{b20}}
 {|-i\omega+ \,u_0\,\nu_0\,k^2|^2}\,{\rm P}_{js}({\bf k}) \, .
\label{prop}
\end{eqnarray}

The model has three triple (interaction) vertices
\begin{eqnarray}
- \bi{v^{\prime}}(\bi{v}\bi{\cdot}\bi{\nabla})\bi{v}&=&v^{\prime}_j
V_{jkl} v_k v_l\,, \\ -
\bi{v^{\prime}}(\bi{b}\bi{\cdot}\bi{\nabla})\bi{b}&=&v^{\prime}_j
V_{jkl} b_k b_l\,, \\
\bi{b^{\prime}}[(\bi{b}\bi{\cdot}\bi{\nabla})\bi{v}-
(\bi{v}\bi{\cdot}\bi{\nabla})\bi{b} ]&=&b^{\prime}_j U_{jkl} b_k
v_l\,,
\end{eqnarray}
where the tensor structure of the vertices in wave-number-frequency
representation are
\begin{equation}
V_{jkl} = i(\delta_{jk} p_l +\delta_{jl} p_k)\,,\,\,\,
U_{jkl}=i(\delta_{jl} p_k -\delta_{jk} p_l),
\end{equation}
where momentum ${\bf p}$ is flowing into the vertex via the
auxiliary fields $\bi{v^{\prime}}$, and $\bi{b^{\prime}}$.

\section{Renormalization}\label{sec3}
\subsection{Divergences of the model}

It can be shown \cite{AdVaHn85} that for any fixed space dimension
$d>2$, the superficial UV divergences can exit only in the following
one-particle irreducible (1PI) Green functions:
$\Gamma^{vv^{\prime}}, \Gamma^{bb^{\prime}}$, and $\Gamma^{v'bb}$.
They lead to local counterterms of the form $\propto
\bi{v^{\prime}}\nabla^2\bi{v}$,
$\propto\bi{b^{\prime}}\nabla^2\bi{b}$, and
$\propto\bi{v^{\prime}}(\bi{b}\bi{\cdot}\bi{\nabla})\bi{b}$ which
are already present in the action (\ref{action}), therefore, the
model is multiplicatively renormalizable (the analytic terms in
$k^2$ proportional to $g_{v20}$, and $g_{v20}$ in (\ref{corelv}),
and (\ref{corelb}) are not needed in this case, and the model can be
formulated without them).

The situation is more complicated in the two dimensional case, where
additional UV divergences appear. They are related to the 1PI Green
functions $\Gamma^{v^{\prime}v^{\prime}}$, and
$\Gamma^{b^{\prime}b^{\prime}}$. In this situation the formulation
of the model without local (analytic in $k^2$) terms cannot give, in
general, multiplicatively renormalizable model because the nonlocal
terms of the action is not renormalized since the divergences
produced by the loop integrals of the diagrams are always local  in
space and time (see, e.g., \cite{Zinn89}). Thus, the simplest way
how to restore the renormalizability of the model (or how to include
the corresponding local counterterms $\propto
\bi{v^{\prime}}\nabla^2\bi{v^{\prime}}$, and
$\propto\bi{b^{\prime}}\nabla^2\bi{b^{\prime}}$ in the
renormalization) is to add corresponding local terms to the force
correlation functions. It is shown explicitly in (\ref{corelv}), and
(\ref{corelb}). In language of classical hydrodynamics the forcing
contribution $\propto k^2$ corresponds to the appearance of large
eddies convected by small and active ones and it is represented by
the local term of $\bi{v^{\prime}}\nabla^2\bi{v^{\prime}}$. In its
analogy the term $\bi{b'}\nabla^2\bi{b'}$ is added to the magnetic
forcing.

Thus, in two dimensions, the model (\ref{action}) is renormalizable
by the standard power-counting rules, and for limits
$\epsilon\rightarrow 0, \ \delta\rightarrow 0$ possesses the
ultraviolet (UV) divergences which are present in five
aforementioned 1PI Green functions. It means that model is
regularized using a combination of analytic and dimensional
regularization with the parameters $\epsilon$, and $\delta=(d-2)/2$.
As a result, the UV divergences appear as poles in $\epsilon$,
$\delta$, and their following combinations: $2\epsilon +\delta$, and
$(a+1)\epsilon + \delta$. The UV  divergences may be removed by
adding needed counterterms to the basic action $S_B$ which is
obtained from the unrenormalized one (\ref{action}) by the
substitution of the renormalized parameters for the bare ones:
$g_{v10}  \rightarrow \mu^{2\epsilon}  g_{v1}$, $g_{v20} \rightarrow
\mu^{-2 \delta} g_{v2}$, $g_{b10}  \rightarrow \mu^{2 a \epsilon }
g_{b1}$, $g_{b20}  \rightarrow \mu^{-2\delta } g_{b2}$, $\nu_0
\rightarrow \nu$, $u_0 \rightarrow u$, where $\mu$ is a
scale-setting parameter having the same canonical dimension as the
wave number.

In what follows, we shall work with, in our case the most
convenient, minimal subtraction (MS) scheme, i.e.,  we are
interesting only in the singular (pole) parts of divergent 1PI Green
functions which are included in the renormalization constants. They
give rise to the counterterms added to the basic action to make the
Green functions of the renormalized model UV finite. In our model,
the counterterms have the form
\begin{eqnarray}\fl
\qquad S_{count}=&\int\,\rmd  x\, \big[ \nu\, \left(1- Z_1\right)
\bi{v}'\nabla^2\, \bi{v}+
     u \nu\,\left(1- Z_2\right) \bi{b}'\nabla^2\, \bi{b}
\nonumber\\
\fl &+\frac{1}{2}\,(\,Z_4-1) u {\nu^3} g_{v2}\,{\mu}^{-2\delta}\,
\bi{v}'{\nabla^2} \bi{v}'  +\frac{1}{2}\, ( Z_5-1 )\,u^2 {\nu^3}
g_{b2}\,{\mu}^{-2\delta}\, \bi{b}'{\nabla^2} \bi{b}'
\nonumber\\
\fl &+(1-Z_3)\,\bi{v}'(\bi{b\cdot\nabla})\bi{b} \big]\,,
\label{counter}
\end{eqnarray}
where the renormalization constants $Z_i, i=1,2,4,5$  renormalizing
the unrenormalized parameters
$e_0=\{g_{v10},g_{v20},g_{b10},g_{b20},\nu_0,u_0\}$, and the
renormalization constant $Z_3$ renormalizing the fields ${\bi b}$,
and ${\bi b^{\prime}}$. They are chosen to cancel the UV divergences
appearing in the Grren functions constructed using the basic action.
The remaining fields $\bi{v^{\prime}}$, and ${\bi v}$ are not
renormalized due to the Galilean invariance of the model
(\ref{action}).

Renormalized Green functions are expressed in terms of the
renormalized pa\-ra\-me\-ters
\begin{eqnarray}
 g_{v1} &= g_{v10}\,{\mu}^{-2\epsilon}\, Z_1^2 Z_2,\qquad &
 g_{v2}  = g_{v20}\,{\mu}^{2\delta}\, Z_1^2 Z_2 Z_4^{-1},
 \nonumber\\
 g_{b1} &= g_{b10}\,{\mu}^{-2\,a\epsilon}\,Z_1 Z_2^{2} Z_3^{-1},\qquad &
 g_{b2}  = g_{b20}\,{\mu}^{2\delta}\, Z_1 Z_2^{2} Z_3^{-1} Z_5^{-1},
 \label{Zkaka1}\\
 \nu &= \nu_0\, Z_{1}^{-1},  \qquad &
   u\ =\ u_0\, Z_2^{-1} Z_1
\nonumber
\end{eqnarray}
appearing in the renormalized action $S^R$ connected with the action
(\ref{action}) by the relation of multiplicative renormalization:
$ S^R \{\bi{e}\} = S \{\bi{e_0}\}$ . The renormalized action  $S^{R}$,
which depends on the renormalized parameters  $e(\mu)$, yields
renormalized Green functions without UV diver\-gences.
The RG is mainly concerned  with the prediction of the asymptotic
behavior of correlation functions expressed in terms of anomalous
dimensions  $\gamma_j$ by  the use of  $\beta$ functions, both defined
via differential relations
\begin{equation}\fl
\qquad \gamma_j=\mu \frac{\partial \ln Z_j}{\partial \mu}\Big|_{e_0}\,,\,\,
\,\,\,
 \beta_g=\mu \frac{\partial g}{\partial\mu }\Big|_{e_0}\,,\qquad\mbox{with}
\,\,\,
 g\equiv\{g_{v1},g_{v2},g_{b1},g_{b2},u\}\,.
 \label{defbeta}
 \end{equation}
These definitions with expressions (\ref{Zkaka1}) yield the $\gamma$-functions
\begin{eqnarray}
\gamma_{gv1}&= -2\gamma_1-\gamma_2\,,
       \qquad& \gamma_{gb1}= -\gamma_1-2\gamma_2+\gamma_3)\,,
 \nonumber\\
\gamma_{gv2}&= -2\gamma_1-\gamma_2+\gamma_4\,,
      \qquad& \gamma_{gb2}= -\gamma_1-2\gamma_2+\gamma_3+\gamma_5\,,
\label{gamy1}\\
\gamma_\nu &= \gamma_1\,, \qquad \gamma_b = {1\over2} \gamma_3\,,
         \qquad &\ \ \gamma_u= -\gamma_1+\gamma_2)
 \nonumber
\end{eqnarray}
and then $\beta$-functions
\begin{eqnarray}
\fl\qquad\ \beta_{gv1}&=g_{v1}\,(-2\epsilon+2\gamma_1+\gamma_2)\,,
    \quad&
 \beta_{gb1}=g_{b1}\,(-2\,a\epsilon+\gamma_1+2\gamma_2-\gamma_3)\,,\,\,
\nonumber \\
\fl\qquad\ \beta_{gv2}&=g_{v2}\,(2\delta+2\gamma_1+\gamma_2-\gamma_4)\,,
    \quad&
 \beta_{gb2}=g_{b2}\,(2\delta+\gamma_1+2\gamma_2-\gamma_3-\gamma_5)
 \label{bety} \\
\fl\qquad\ \ \ \beta_u&= u\,(\gamma_1-\gamma_2).
\nonumber
\end{eqnarray}

\subsection{RG equations}

Correlation functions of the fields are expressed in terms of
scaling functions of the variable $s=(k/\mu)$,
$s\in\langle0,1\rangle$. Then the asymptotic behaviour and the
universality of MHD statistics stem from the existence of a stable
IR fixed point. The continuous RG transformation is an operation
linking a sequence of invariant parameters ${\overline g}(s)$
determined by the Gell-Mann-Low equations
\begin{equation}\fl
\qquad  \frac{d\overline g(s)}{d\ln s}= \beta_g\left({\overline g}(s)\right)\,\,\,
\,\mbox{with the abbrevation}\,\,\,{\overline g}\equiv \{ {\overline g_{v1}},
 {\overline g_{v2}}, {\overline g_{b1}}, {\overline g_{b2}}, {\overline u} \}\,,
\label{Gell}
\end{equation}
where the scaling variable $s$ parameterizes RG flow with the
initial conditions ${\overline g}|_{s=1}\equiv g$ (the critical
behaviour corresponds to IR limit $s\rightarrow 0$). The expression
of the $\beta({\overline g}(s))$ function is known in the framework
of the $\delta$, $\epsilon$ expansion  (see (\ref{gama}) and also
(\ref{bety})). The fixed point $g^{\ast}(s\rightarrow 0)$ satisfies
a system of equations $\beta_{g}(g^{\ast})=0, $ while a IR stable
fixed point, weakly dependent on initial conditions, is defined by
positive definiteness of the real part of  the matrix   $\Omega=
(\partial \beta_g/\partial g)|_{g^{\ast}}$ (the matrix of the first
derivatives taken at the fixed point). In other words, a fixed point
is stable if all the trajectories $g(s)$ in its vicinity approach
the value of the fixed point.

The initial conditions ${\overline g}|_{s\rightarrow 1}=g$ of the
equations (\ref{Gell}), dictated by a micromodel, are insufficient
since our aim is the large-scale limit of statistical theory, where
$g^{\ast}\equiv {\overline g}|_{s\rightarrow 0}$. As was mentioned
already, the RG fixed point is defined by the equation
\begin{equation}\beta(\,g^{\ast}\,)= 0\,.\label{beta01}\end{equation}
For ${\overline g}(s)$ close to $g^{\ast}$ we obtain a system
of linearized equations
\begin{equation}
\left( I\,s\,\frac{\mbox{d}}{\mbox{d} s} - \Omega\,
\right) \,({\overline g} - g^{\ast})\,= 0,
\end{equation}
where $I$ is $(5\times 5)$ unit matrix. Solutions of this system
behave like ${\overline g}=g^{\ast}+{\cal O}( s^{{\xi}_j}) $
if $s\rightarrow 0$.
The exponents $\xi_j$ are the elements of the diagonalized
matrix $\Omega^{diag} =
(\,\xi_1,\,\xi_2,\,\xi_3,\,\xi_4, \xi_5 \,)$
 and can be obtained as roots of the characteristic polynomial
 ${\mbox{\small Det}}(\Omega - \xi I).$
 The positive defineteness of $\Omega$ represented by the conditions
 $\mbox{Re}_j(\xi)\geq 0, j=1,2,...5$ is the test of the IR
 asymptotical stability of discussed theory.

\subsection{One-loop order calculation}

In the standard MS scheme \cite{Hoft73} the renormalization constants have the general form
\begin{equation}
Z_i=1-F_i P^{\delta,\epsilon},
\end{equation}
where the terms $P^{\delta,\epsilon}$ are given by the linear combinations of the poles
and the amplitudes $F_i$ are some functions of  $g_{v1},g_{v2},g_{b1},g_{b2}$, and $u$,
but are independent of $\delta$ and $\epsilon$.  The amplitudes $F_i=F_i^{(1)}
F_i^{(2)}$ are a product of two multipliers $F_i^{(1)}, F^{(2)}_i.$
One of them, say, $F_i^{(1)}$ is a multiplier originating from the
divergent part of the Feynman diagrams, and the second one,
$F_i^{(2)}$ is connected only with the tensor nature of the
diagrams (see discussion in \cite{HnJoJuSt01} for details).

It can be explained by the following simple example \cite{HnJoJuSt01}
(the example is taken from a problem with anisotropy, i.e., where another
arbitrary unit vector ${\bf n}$ exists but the conclusions are the same).
Consider a UV-divergent integral
$$ I({\bf k}, {\bf n}) \equiv n_i n_j k_l k_m\int d^d {\bf q}
\frac{1}{(q^2+m^2)^{1+2\delta}} (\frac{q_iq_jq_lq_m}{q^4}
-\frac{\delta_{ij}q_lq_m+\delta_{il}q_jq_m+\delta_{jl}q_iq_m}{3q^2})$$
(summations over repeated indices are implied) where $m$ is an
infrared mass. It can be simplified in the following way: $$ I({\bf
k}, {\bf n}) \equiv n_i n_j k_l k_m S_{ijlm}\int_0^{\infty} d q^2
\frac{q^{2\delta}}{2(q^2+m^2)^{1+2\delta}},$$ where
$$S_{ijlm}=\frac{S_d}{d(d+2)}(\delta_{ij}\delta_{lm}+\delta_{il}\delta_{jm}+
\delta_{im}\delta_{jl}-\frac{(d+2)}{3}(\delta_{ij}\delta_{lm}+
\delta_{il}\delta_{jm}+\delta_{im}\delta_{jl})),$$
$$\int_0^{\infty} d q^2
\frac{q^{2\delta}}{2(q^2+m^2)^{1+2\delta}}=\frac{\Gamma{(\delta
+1)}\Gamma{(\delta)}}{2 m^{2\delta}\Gamma(2\delta+1)}, $$ and
$S_d=2\pi^{d/2}/\Gamma(d/2)$ is the surface of unit the
d-dimensional sphere. The purely UV divergent part manifests itself
as the pole in $2\delta=d-2$; therefore, we find $$\mbox{UV div.
part of}\,\,\,\, I= \frac{1}{2\delta}(F^{(2)}_1 k^2+F^{(2)}_2 ({\bf
n k})^2),$$ where $F^{(2)}_1=F^{(2)}_2/2 = (1-d)S_d/3d(d+2)$
($F_1^{(1)}=F_2^{(1)}=1$).
It has to be mentioned that in spite of the above simple example in our
calculation we shall introduce the needed IR regularization by restriction
on interval of integrations.

In the standard MS scheme one puts $d=2$ in $F^{(2)}_1,F^{(2)}_2$,
therefore the d-dependence of these multipliers is ignored. As was
discussed in \cite{HnJoJuSt01},  for the theories with vector fields
and, consequently, with tensor diagrams, where the sign of values of
fixed points and/or their stability depend on the dimension $d$, the
procedure, which eliminates the dependence of multipliers of the
type $F^{(2)}_1,F^{(2)}_2$ on $d,$ is not completely correct because
one is not able to control the stability of the fixed point when
$d=3.$ Therefore, in \cite{HnJoJuSt01} it was proposed to slightly
modify the MS scheme in such a way to keep the d-dependence of $F$
in renormalization constants $Z_i$. Then the subsequent calculations
of the RG functions ($\beta-$ functions and anomalous dimensions
$\gamma_i$) allow one to arrive at the results which are in
qualitative agreement with the results obtained in the framework of
the simple analytical regularization scheme, i.e., one is able to
obtain the fixed point which is not stable for $d=2$, but whose
stability is restored for a borderline dimension $2<d_c<3$. In what
follows, it will be shown that it is really our case, thus we shall
apply this modified MS scheme in our calculations.

Now we can return and continue with RG analysis. Using the RG
routine the anomalous dimensions
$\gamma_j(g_{v1},g_{v2},g_{b1},g_{b2})$ can be extracted from
one-loop diagrams.  Thus, the extraction of the UV-divergent parts
from one-loop diagrams gives $Z$-constants in the form
\begin{eqnarray}
\fl\quad Z_1&=&1+\frac{S_d}{(2\pi)^d} \left[ \,u\,\lambda_5 \left(
 \frac{ g_{v2}}{2\delta}-\frac{ g_{v1}}{2\epsilon}\right) + \lambda_6 \left(
 \frac{ g_{b2}}{2\delta}-\frac{ g_{b1}}{2 a\epsilon}\right) \right]\,,
\nonumber\\
\fl\quad Z_2&=&1+\frac{S_d}{(2\pi)^d (u+1)} \left[\lambda_1 \left(
 \frac{ g_{v2}}{2\delta}-\frac{ g_{v1}}{2\epsilon}\right) +
 \lambda_3 \left(\frac{ g_{b2}}{2\delta}-\frac{ g_{b1}}{2 a \epsilon} \right)\right]\,,
\nonumber\\
\fl\quad Z_3&=&1+\frac{S_d}{(2\pi)^d} \,\lambda_7 \left(
 \frac{ g_{v1}}{2\epsilon}-\frac{ g_{v2}}{2\delta} -
 \frac{ g_{b1}}{2a\epsilon}+\frac{ g_{b2}}{2\delta}\right)\,,
\label{zetka}\\
\fl\quad Z_4&=&1+\frac{S_d}{(2\pi)^d} \frac{\lambda_4}{g_{v2}}\left(
\frac{u g_{v1}^2}{2\delta+4\epsilon}+\frac{2u g_{v1}g_{v2}}{2\epsilon}
  -\frac{u g_{v2}^2}{2\delta}  +
 \frac{g_{b1}^2}{2\delta+4a\epsilon}+\frac{2 g_{b1}g_{b2}}{2a\epsilon}
  -\frac{g_{b2}^2}{2\delta}  \right)\,,
\nonumber\\
\fl\quad Z_5&=&1+\frac{S_d}{(2\pi)^d} \frac{\lambda_2}{(u+1)g_{b2}} \left(
\frac{g_{v1}\,g_{b1}}{2\delta+2\epsilon(1+a)}+
\frac{g_{v1}\,g_{b2}}{2\epsilon}+\frac{g_{v2}\,g_{b1}}{2 a\epsilon}
 -\frac{g_{v2}\,g_{b2}}{2\delta}  \right) \,,
\nonumber
\end{eqnarray}
and, in consequence, the lowest order $\gamma$-functions are
\begin{eqnarray}
\gamma_1&=&\widetilde{S_d}\,\left( u\,\lambda_5\,g_v + \lambda_6\,g_b \right)\,,
\quad\ \,\gamma_2=\widetilde{S_d}\,\frac{(\lambda_1\,g_v + \lambda_3\,g_b)}{u+1}\,,
\nonumber\\
\gamma_3&=&\widetilde{S_d}\,\lambda_7\, ( -g_v + g_b )\,,
\qquad\quad\,\gamma_4=\widetilde{S_d}\,\frac{\lambda_4}{g_{v2}}( u\,g_v^2 + g_b^2 )\,,
\label{gama}\\
\gamma_5&=&\widetilde{S_d}\,\frac{\lambda_2}{(1+u)}\frac{g_v\ g_b}{g_{b2}}\,,
\nonumber
\end{eqnarray}
where $\widetilde{S_d}=S_d/(2\pi)^d$, $S_d$ denote $d$-dimensional sphere
$S_d=2\pi^{d/2}/\Gamma(d/2)$, $g_v\equiv g_{v1}+g_{v2}$, $g_b\equiv g_{b1}+g_{b2}$,
and $\lambda$-coefficients depend only on dimension $d$:
\begin{eqnarray}\fl\qquad
\lambda_1&=\frac{d-1}{2d}\,, \quad\quad\ \lambda_2=\frac{d-2}{2d}\,,  \quad &
\lambda_3=\frac{d-3}{2d}\,,  \quad\ \    \lambda_4=\frac{d^2-2}{4d(d+2)}\,,
\nonumber \\
\fl\qquad
\lambda_5&=\frac{d-1}{4(d+2)}\,, \quad \lambda_6=\frac{d^2+d-4}{4d(d+2)}\,,\quad &
\lambda_7=\frac{1}{d(d+2)}\,.
\label{lambdy}
\end{eqnarray}
Substituting (\ref{gama}) into $\beta$-functions (\ref{bety}) one can
obtains $\beta$-functions in the one-loop order approximation.
Note that in two dimensions the $\gamma$-functions are
\begin{eqnarray}\fl\qquad
\gamma_1^{(2)}&=&{1\over32\pi} \left( u\,g_v + g_b \right)\,,\quad\,
\gamma_2^{(2)} = {1\over8\pi} \frac{(g_v - g_b)}{(u+1)}\,,   \quad\,
\gamma_3^{(2)} = {1\over16\pi}\, (-g_v + g_b)\,,
\nonumber\\ \fl\qquad
\gamma_4^{(2)}&=&{1\over32\pi} \frac{( u\,g_v^2 + g_b^2 )}{g_{v2}}\,,\quad\,
\gamma_5^{(2)} = 0
\label{gamaD2}
\end{eqnarray}
and, in correspondence with \cite{HnHoJu01} $Z_5=1$, which is a specific property
of the two-dimensional MHD turbulence because there are no UV divergences in
the 1PI Green's function $\Gamma^{b'b'}$ in the one-loop approximation.  Here we
emphasize that in general case of $d$ dimensions $\gamma_5\neq 0$ and $Z_5\neq 1$.

\section{Fixed points}\label{sec4}
\subsection{Case of passive vector admixture}

Here we briefly consider the case when the magnetic field can be
treated as a passive vector field in the developed HD turbulence.
Notation the "passive" magnetic field means that the Lorentz force
acting on conductive fluid can be neglected at large spatial scales,
thus, the Lorentzian term $(\bf{b\cdot\nabla})\bf{b}$ in the
Navier-Stokes equation can be omitted. Just then the vertex function
$\Gamma^{v'bb}$ is finite and the term containing $Z_3$ in
$S_{count}$ does not exists. Therefore, the magnetic field is not
renormalized and $\gamma_3=0$. Furthermore, some diagrams of
$\Gamma^{v'v}, \Gamma^{v'v'}, \Gamma^{b'b}$ containing the vertex
$\Gamma^{v'bb}$ can be omitted and $Z$-constants as well as
$\gamma$-functions are reduced. Resulting $\gamma$-functions take
the form
\begin{eqnarray}
\gamma_1 &=& \widetilde{S_d}\, u\,\lambda_5\,g_v \,,\quad\ \,
\gamma_2\ =\ \widetilde{S_d}\,\lambda_1\,\frac{g_v}{u+1}\,,
\nonumber\\
\gamma_4 &=& \widetilde{S_d}\,\lambda_4\,\frac{u}{g_{v2}}\,g_v^2 \,,\quad\,
\gamma_5\ =\ \widetilde{S_d}\,\frac{\lambda_2}{(1+u)}\frac{g_v\ g_b}{g_{b2}} \, .
\label{gamabez}
\end{eqnarray}
Substituting of $\gamma$-functions (\ref{gamabez}) and $\gamma_3=0$ into
$\beta$-equations (\ref{bety}) one obtains a system of fourth nonlinear equations
$\beta_{gv1}=\beta_{gv2}=\beta_{gb1}=\beta_{gb2}=0$ for $g_i$ and
one equation $\beta_u=0$ for $u$. The last one gives $u^\ast=0$, or, nonzero
universal inverse Prandtl number,
\begin{equation}
 u^{\ast}={1\over2}\left( \sqrt{\frac{16+9 d}{d}}-1\right) \,.
\label{uu}
\end{equation}
In the first case of $u^\ast=0$ one obtains only two fixed points (with zeroth
$g_{b1}^\ast, g_{b2}^\ast$):\\
1. $g_{v1}^\ast=0, g_{v2}^\ast=-2\delta/\lambda_1 \widetilde{S_d}\ $
which is non-physical (negative), and, \\
2. $g_{v1}^\ast=2\epsilon/\lambda_1 \widetilde{S_d}, g_{v2}^\ast=0\ $ which in unstable.

Let $u$ is given by (\ref{uu}). Then apart from the Gaussian fixed
point $g_{v1}^\ast=g_{v2}^\ast=g_{b1}^\ast=g_{b2}^\ast=0$, with no
fluctuation effect on the large-scale asymptotics, there are
following fixed points with $g_{b2}^\ast=0$:
\begin{eqnarray*}
\fl\qquad  (1^\ast)\qquad  g_{v1}^\ast=0,\quad
    g_{v2}^\ast=-\frac{2(d-2)d^2(u^\ast+1)}{2d^2-3d+2}\widetilde{S_d}^{-1},\quad
    g_{b1}^\ast=0;
\\
\fl\qquad  (2^\ast)\qquad
   g_{v1}^\ast=\frac{4\epsilon d(u^\ast+1)}{3(d-1)}\widetilde{S_d}^{-1},\quad
    g_{v2}^\ast=0,\quad g_{b1}^\ast=0;
\\
\fl\qquad  (3^\ast)\qquad
g_{v1}^\ast=\frac{4\epsilon(3d^3+d^2(4\epsilon-9)-6d(\epsilon-1)+4\epsilon)(u^\ast+1)}
                      {9(d+2\epsilon-2)(d-1)^2}\widetilde{S_d}^{-1},
\\
\fl\qquad\qquad\qquad  g_{v2}^\ast=\frac{8\epsilon^2(d^2-2)(u^\ast+1)}
              {9(d+2\epsilon-2)(d-1)^2}\widetilde{S_d}^{-1},\quad g_{b1}^\ast=0.
\end{eqnarray*}
Next three fixed points are the same as the last $(1^\ast)$--$(3^\ast)$
with different $g_{b2}^\ast$:\\
$(1a^\ast)$ $\ \ g_{b2}^\ast=(d^2-2)/d(d-2)$; \\
$(2a^\ast)\equiv(3a^\ast)$  $\ \ g_{b2}^\ast=3(d-1)(d+2\epsilon-2/2(d-2)\epsilon$.\\
The points $(2a^\ast)$ and $(3a^\ast)$ have the same $g_{b2}^\ast$ because
$g_{v1}^\ast$ of the point $(2^\ast)$ is equal to the sum $(g_{v1}^\ast + g_{v2}^\ast)$
of the point $(3^\ast)$. Note that $g_{b2}^\ast$ has discontinuity at $d=2$.

The "thermal" point $(1^\ast)$ is generated by short-range correlations of the random
force \cite{HnHoJu01} and has negative $g_{v2}^\ast$. The second fixed point $(2^\ast)$
is unstable.  The physical meaning has the third "kinetic" point $(3^\ast)$ whose
parameters $\{g_1, g_2, u\}$ dependence on the dimension $d$ is shown in figure~1.
for physical value of $\epsilon=2$.
%
%   <<<<<   Fig.1  <<<<<  Kinet.bod  G1,G2,U(d)   <<<<<<<<<<<<<<<<<<<<<<<<<<<<<<<<
 \begin{figure}[ht] % [ht] here, [t] top, [e] end
 \vspace*{-3mm}
\begin{center}
\includegraphics[angle=-90,width=90mm]{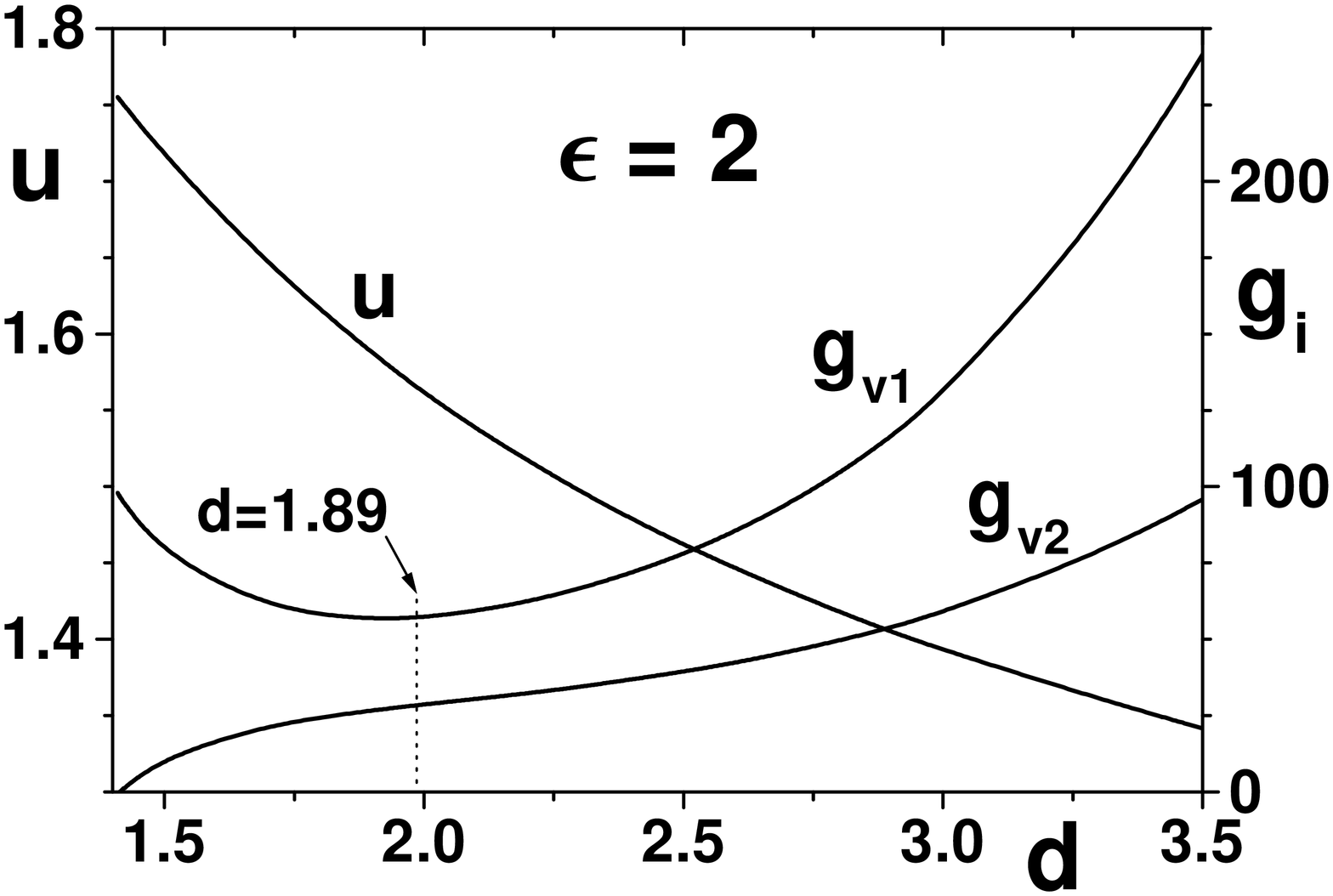}
\vspace{-3mm}
\caption{Dependence of the parameters $\{g_{v1}, g_{v2}, u\}$ \\
    on the dimension $d$ for $\epsilon=2$ at the kinetic fixed point (\ref{fixed2bez}).}
\end{center} \label{f1}
\end{figure} %<<<<<<<<<<<<<<<<<<<<<<<<<<<<<<<<<<<<<<<<<<<<<<<<<<<<<<<<<<<<<<<<<<<<
\\
Setting $\epsilon=2$ and $u^\ast$ from (\ref{uu}) one obtains
\begin{equation}\fl\ \
g_{v1}^\ast =  \frac{(2\pi)^d}{S_d}
  \frac{8(u^\ast+1)(3 d^3-d^2-6 d+8)}{9(d-1)^2(d+2)} \,,\quad
g_{v2}^\ast =  \frac{(2\pi)^d}{S_d}
              \frac{32(u^\ast+1)(d^2-2)}{9(d-1)^2(d+2)}\,.
\label{fixed2bez}
\end{equation}
In this case the sum of $g_{v1}^\ast+g_{v2}^\ast\equiv g_{v}^\ast =
(2\pi)^d\, 8 d (u^\ast+1)/3(d-1){S_d}$.
Detailed numerical calculations have shown that the region of stability
of this point is limited by the value of parameter $\ a<1\ $ and this
limiting value does not depend on the dimension $\ d$. This stable region
is denoted as region $A$ in figure~\ref{fig2}.
%
%   <<<<<   Fig.2  <<<<<  Kinet.bod = Oblast stability   <<<<<<<<<<<<<<<<<<<<<<<
 \begin{figure}[ht]  % [ht] here, [t] top, [e] end
 \vspace*{-3mm}
\begin{center}
\includegraphics[angle=-90,width=90mm]{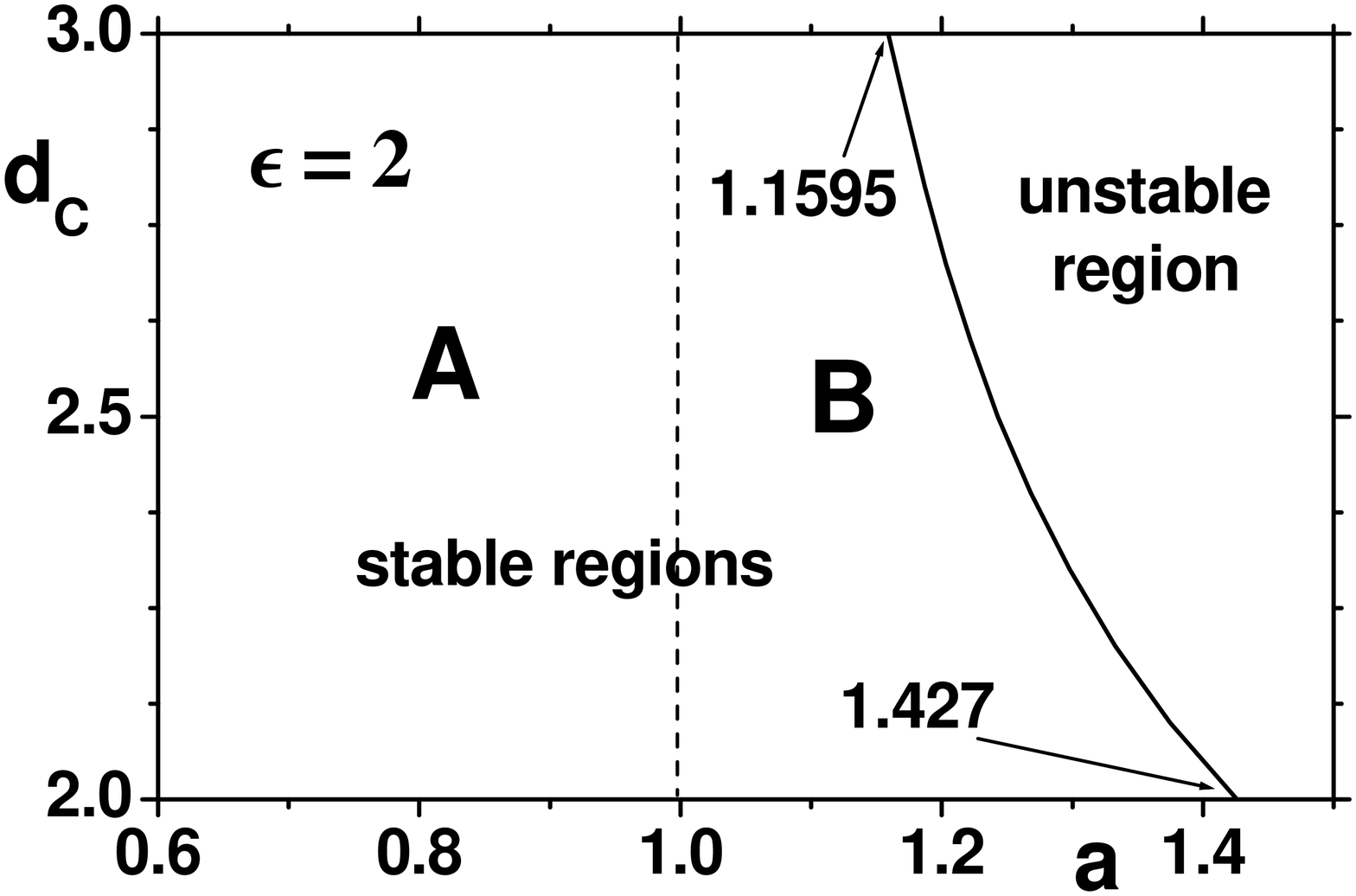}
\end{center}
\vspace{-6mm}
\caption{Stability regions of kinetic point and the critical dimension $d_c$ \\
      dependence on parameter $a$. The region $A$ spreads down to $a=0$.}
 \label{fig2}
 \end{figure}

\subsection{Case of active vector admixture}

In the full self-consistent system, the RG equations yield besides
the known fixed point in the kinetic regime also the nontrivial
magnetic fixed point. If the both are stable in the same region of
parameters then the choice between two possible critical regimes
will depend on initial conditions for RG equations, i.e. critical
behavior of the system is non universal.

\subsubsection{Kinetic fixed point.}

\hspace*{0mm}
The nontrivial stable kinetic fixed point of RG equations has been found to be the same
as in the previous case of passive magnetic field admixture because the $\beta$-functions
$\beta_{gv1}, \beta_{gv2}$ are the same for zero $g_{b1}, g_{b2}$. Only difference was
found in the stability region in dependence on parameter $a$: the stable region is
enlarged by new region $B$ unlike the case of passive magnetic field admixture, see
figure~\ref{fig2}.  The critical dimension $d_c$ continuously decreases from $3$
to $2$ in dependence on value of parameter $a$ from the interval
$\langle 1.1595, 1.427\rangle$.
It confirms the results of \cite{HnHoJu01} that the stability of kinetic scaling regime
is strongly affected by the behavior of magnetic fluctuations.

\begin{figure}[ht]  % [ht] here, [t] top, [e] end
\vspace*{-3mm}
 \begin{center}
  \begin{flushleft}  \includegraphics[angle=-90,width=67mm]{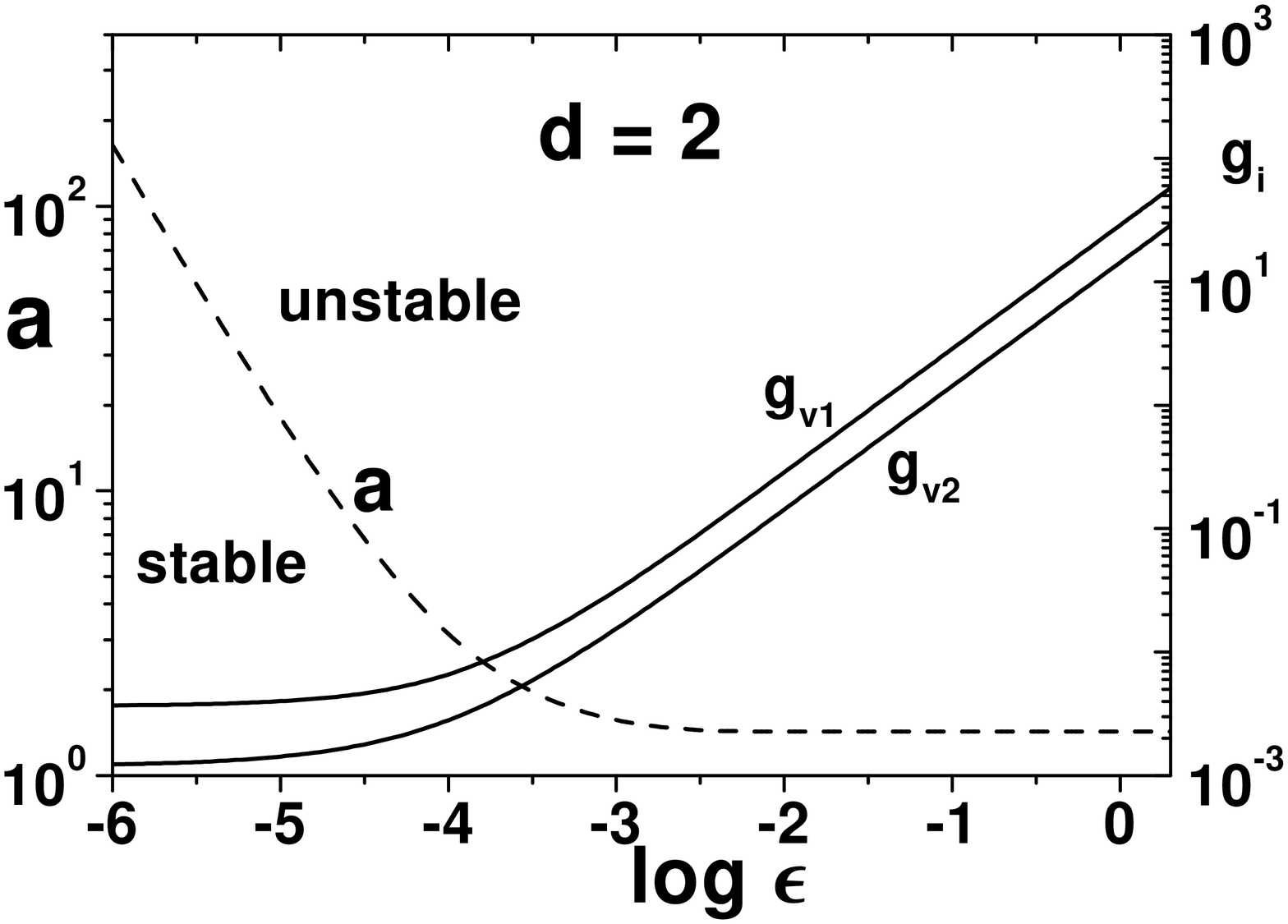}
  \end{flushleft}    \vspace{-50mm}
  \begin{flushright} \includegraphics[angle=-90,width=67mm]{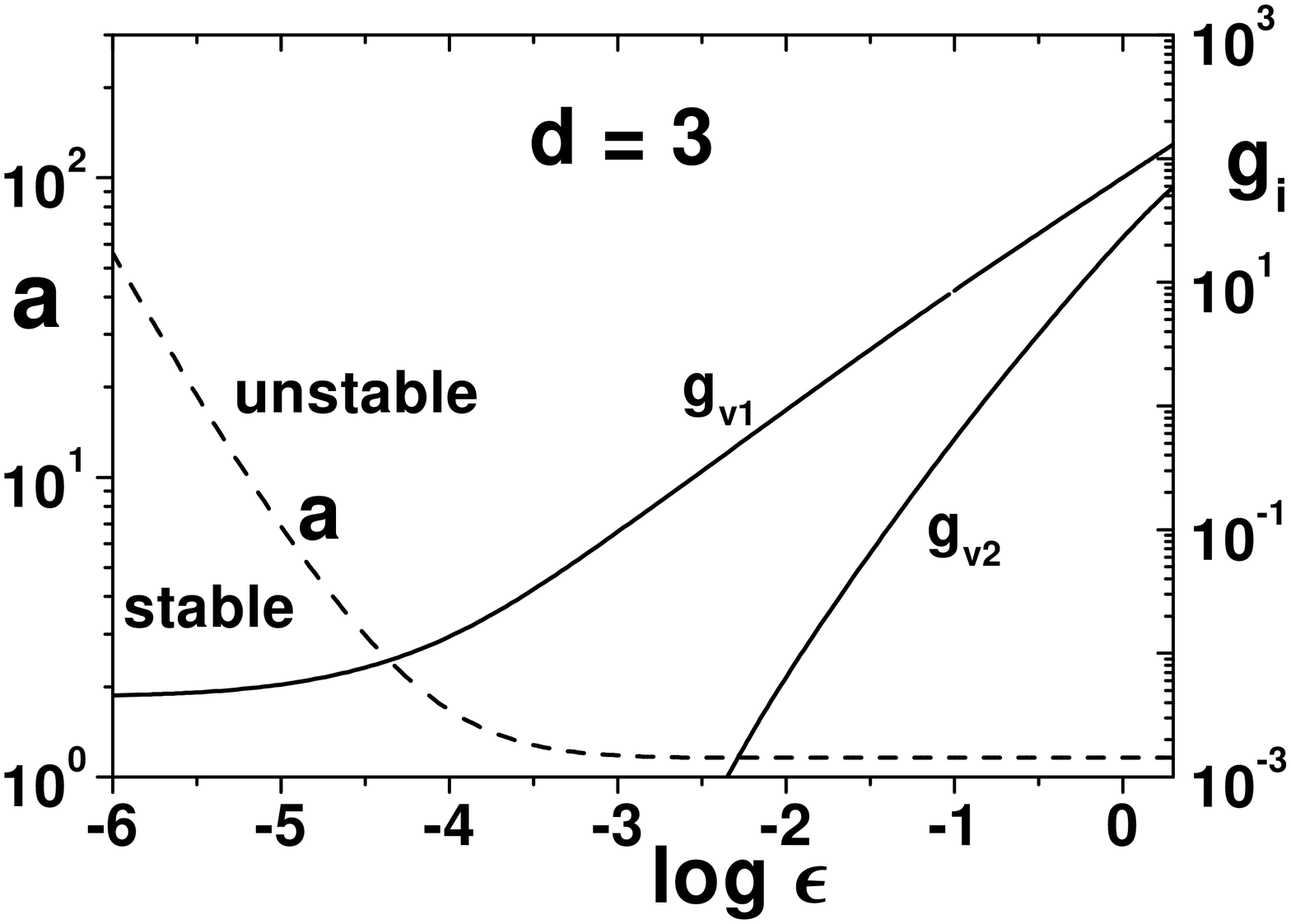}
  \end{flushright}
 \end{center}
\vspace{-4mm}
\caption{Dependence of the parameters $\{g_{v1}, g_{v2}, u\}$ on value of $\epsilon$ \\
         for $2$- and $3$-dimensions at the kinetic fixed point in the general case. \\
         Dashed line shows the critical value of $a$ at the stability region limit.}
\label{fig3ab}
\end{figure}

Figure~\ref{fig3ab} shows values of the charges $g_{v1}, g_{v2}$ which continuously
depend on value of nonzero $\epsilon\leq 2$, for two special case of $d$ equal
to $2$ and $3$. The right axle corresponds to the physical value of $\epsilon=2$.
While the both charges remain nonzero (positive) for $d=2$, in three dimensions
one of them, $g_{v2}$ rapidly decreases for $\epsilon\rightarrow 0$.
The stable as well as unstable regions depends on parameter $a$ and the critical
value of $a$ remains the same for $\epsilon=2$ following from figure~\ref{fig2},
or greater for $\epsilon<2$ (the critical $a$ increases for $\epsilon\rightarrow 0$).
%
%    <<<<<  Fig.3.ab  <<<<<  Kin.bod = zav. od epsilon (d=2; 3)  <<<<<<<<<<<<<<<<<

\subsubsection{Magnetic fixed point.}

\hspace*{0mm} We have shown in (\ref{gamaD2}) that in two dimensions
the function $\gamma_5$ vanishes and then both functions
$\beta_{gb1}$ and $\beta_{gb2}$ contain the same linear combination
of $\gamma$ functions. Thus, at least one of the magnetic charges
($g_{b1}, g_{b2})$ must be zero in fixed point. But in the other
dimensions this restriction does not take place.

Here we restrict ourselves only by finding nontrivial magnetic fixed point.
In \cite{HnHoJu01} there was mentioned that it is characterized by zero
$g_{v1}^\ast$ and $u^\ast$. Therefore, the set of five equations of zero
$\beta$-functions (\ref{bety}) is reduced to three equations. Applying  $\ g_{v1}=u=0\ $
in (\ref{bety}), (\ref{gama}) and (\ref{lambdy}) one obtains the set
\begin{eqnarray}
a_1g_{v2} +a_2g_{v2}^2 +a_3g_{v2}g_b -a_4g_b^2 &=& 0 \ ,
\nonumber\\
-A_0 +a_5g_{v2} +a_6g_b &=& 0 \ ,
\nonumber\\
a_1g_{b2} +a_5g_{v2}g_{b2} +a_6g_{b2}g_b -a_7g_{v2}g_b &=& 0 \ ,
%\nonumber\\
%-g_{b1} +g_{b2} +g_b &=& 0 \ ,
\label{betared}
\end{eqnarray}
where
\begin{eqnarray}
A_0 &=& \frac{2a\epsilon}{S_d}\ ,\qquad\quad\ a_1=\frac{2(d-2)}{S_d}\ ,
\quad\quad  a_2=\frac{(d-1)}{2d}\ ,
\nonumber \\
a_3 &=& \frac{(d^2-5)}{d(d+2)}\ ,\quad\  a_4=\frac{(d^2-2)}{4d(d+2)}\ ,
\quad\  a_5=\frac{(d^2+d-1)}{d(d+2)} ,
\nonumber \\
a_6 &=& \frac{(5d^2-3d-32)}{4d(d+2)}\ ,\qquad\qquad\qquad\quad
a_7=\frac{(d-2)}{2d} \ . \label{acoef}
\end{eqnarray}
Positive coefficients $a_1, a_7$ vanishes at $d=2$, $a_3$ and $a_6$ are positive
for $d>2.236$ and $d>2.848$, respectively.
%
%    <<<<<  Fig.4  <<<<<  Mag.bod = Oblast stability  <<<<<<<<<<<<<<<<<<<<
 \begin{figure}[ht]  % [ht] here, [t] top, [e] end
 \vspace*{-4mm}
\begin{center}
\includegraphics[angle=-90,width=90mm]{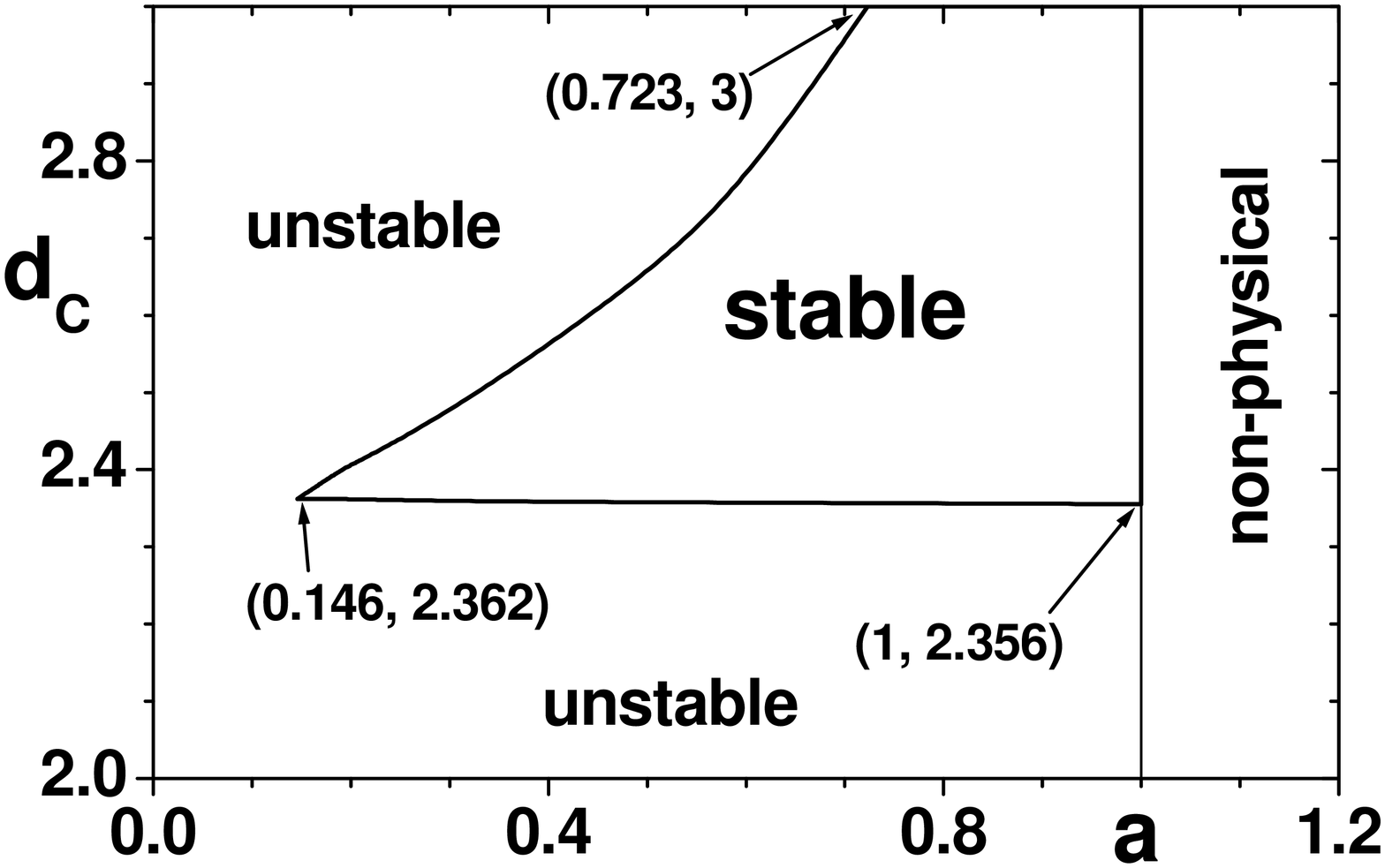}
\end{center}
\vspace{-6mm}
\caption{The stability region of the magnetic fixed point \\
         in the plane of $\{d,a\}$ for the physical value of $\ \epsilon=2$.}
\label{fig4}
\end{figure}
The set (\ref{betared}) can be analytically solved with respect to $g_{v2}, g_{b1},
g_{b2}$. Because all $g_i$ must be positive, the system (\ref{betared}) with
$\ g_{v1}=u=0\ $ gives the only solution,
\begin{eqnarray}
g_{v2} &=& \frac{A_0 - a_6 g_b}{a_5} \,, \qquad
g_{b1}  =  g_b -\frac{g_b(a_6g_b-A_0)-a_5a_7}{a_5(a_1+2a_6g_b-A_0)} \ ,
\nonumber\\
g_{b2} &=& \frac{g_b(a_6g_b-A_0)-a_5a_7}{a_5(a_1+2a_6g_b-A_0)} \ ,
\label{solut}
\end{eqnarray}
where
\begin{eqnarray}
g_b = \frac{-a_1a_5a_6 +a_3a_5A_0 -2a_2a_6A_0 +a_5\sqrt{D}}
             {2(a_4a_5^2 +a_3a_5a_6 -a_2a_6^2)} \,,
\label{denominator}\\
D =a_1^2a_6^2 +4a_1a_4a_5A_0 +2a_1a_3a_6A_0 +a_3^2A_0^2 +4a_2a_4A_0^2\,.
\nonumber
\end{eqnarray}
Note that the parameters $a$ and $\epsilon$ appears in the solution
only as the product $a\epsilon$ in $A_0$. The physical value is
restricted by the inequality $a\epsilon\leq2$. Denominator in the
expression (\ref{denominator}) for $g_b$ is zero at $d_0=2.2628$
(and it is positive for $d>d_0$), therefore, at $d_0$ we can expect
discontinuity and/or divergence. Numerical analysis of the
expressions (\ref{solut}) shows that all $g_i$ have a discontinuity
at $d_0$, and, a physical solution cannot exist for any $a,
\epsilon\ $ if $d\leq d_0$. The stability region of the magnetic
fixed point and the corresponding critical dimension $d_c$ was
determined numerically and it is shown in figure~\ref{fig4}.
%
%    <<<<<  Fig.5.abc <<<<<  Mag.bod = zav. Gv2,Gb1,Gb2(d)  <<<<<<<<<<<<<<<<<<
 \begin{figure}[ht]  % [ht] here, [t] top, [e] end
 \vspace*{-7mm}
\begin{center}
 \includegraphics[width=90mm]{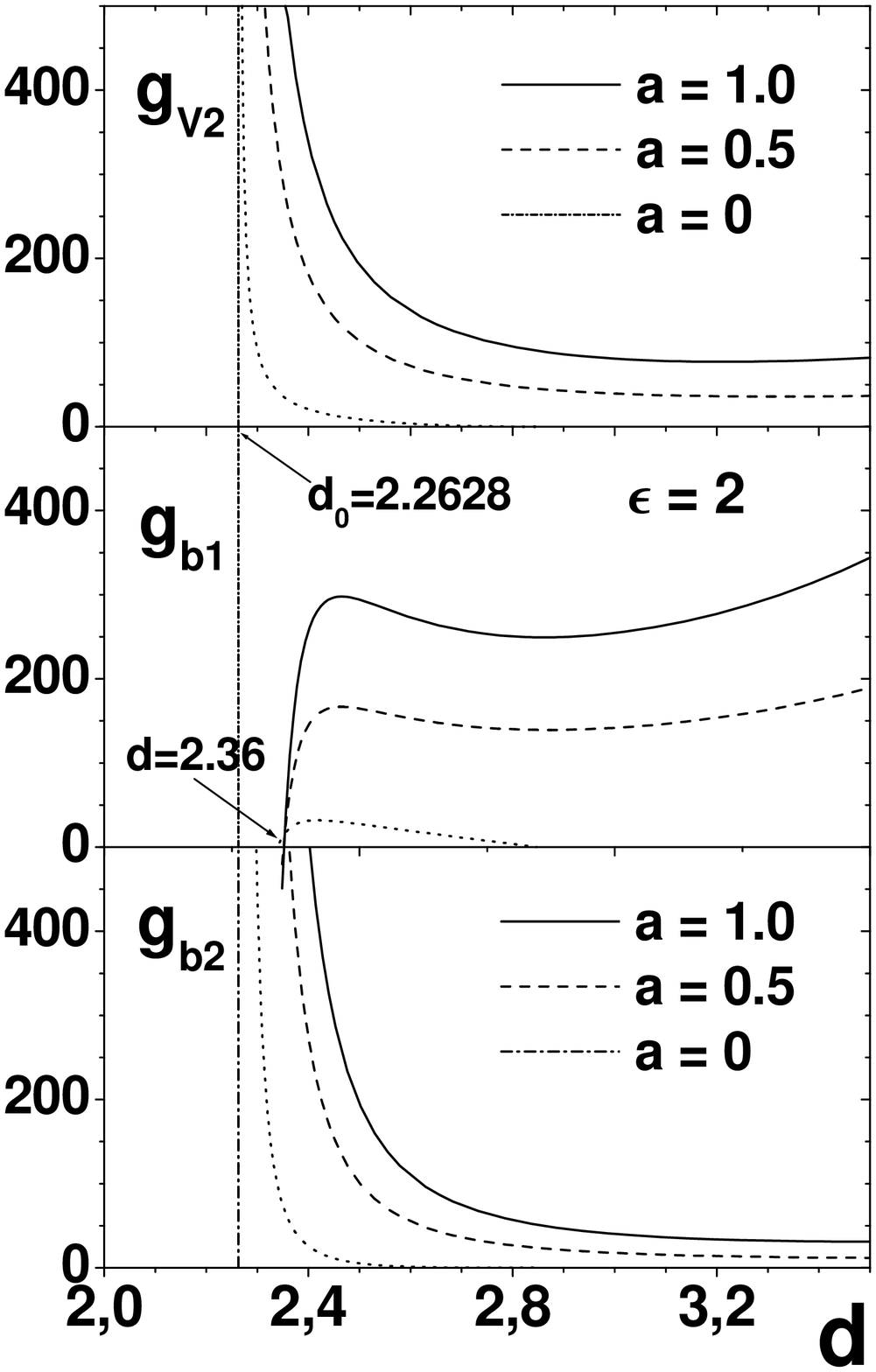}
\end{center}
\vspace{-13mm}
\caption{Dependence of the parameters $\{g_{v2}, g_{b1}, g_{b2}\}$ on the dimension $d$ \\
         for $\epsilon=2$ at the magnetic fixed point (\ref{solut}) for $a=1, 0.5, 0$.
         All $g_i$ have \\    discontinuity at $d=2.2628$ (chained vertical line).}
 \label{fig5}
 \end{figure}
\\
Figure~\ref{fig5} demonstrates the charges $g_{v2}, g_{b1}, g_{b2}$
dependence on dimension $d$ for several values of parameter $a$.
First, we have found that $g_{v2}, g_{b2}$ tend to infinity at limit
value $d_0$. For increasing dimension $d$ from $2$ up to $d_0$ the
charge $g_{b2}$ increase from a small positive value up to infinity
at $d_0$ and, therefore, $g_{v2}$ decrease here from a small
negative value to minus infinity at $d_0$ (because $g_{v2}\propto
-a_6 g_b$ and both $a_6$ and $g_b$ are negative in this dimensions).
The charge $g_{b1}$ rapidly decrease to zero at $d=2.352$ for
decreasing $d$ and continue to minus infinity at $d_0$. These
limiting value are in correspondence with numerical calculation of
the stability region - the system losses stability for the critical
dimension $d_c$ lower than approximately $2.36$ for arbitrary
parameter $a$.
% ----------------------------------------------------------------------

\section{Discussion and conclusions}\label{sec5}

In this paper we revised the calculations of stability ranges of
developed magnetohydrodynamic turbulence in the frame of double
expansion scheme. The modified standard minimal subtraction scheme
\cite{HnJoJuSt01} has been used in the dimension region of $ d\geq2
$ up to $ d=3 $ in both cases of the magnetic field treated as a
passive as well as active vector admixture.
We confirm existence of the known "kinetic" fixed point (corresponding to
Kolmogorov scaling regime) what is the same
in the both considered cases and only difference is in the stability
region: the critical dimension $d_c$ is achieved for a slightly
higher value of $a$-parameter of a magnetic forcing in the case of
active magnetic field. Limit value of the inverse Prandtl number at
$d=3$ restores the value of $ u=1.393 $ which is known from usual
$\epsilon$-expansion, and it fluently rises to $u=1.562$ at $d=2$,
(figure~1).

It was believed earlier that in the double expansion being defined for
the space dimension to be closed to two the results obtained in two dimensions can not be
applicable to opposite dimension interval end closed to three. Here we have showed that
the double expansion in exact $d$-dimensional formulation gives some critical dimension
$d_c$ above which the scaling regime is governed by the competition of the stable kinetic
and magnetic fixed points which exists in three dimensions.

  A new nontrivial results of the present paper is connected with derivation of the exact
analytical expression for the nontrivial "magnetic" stable fixed point with $u=g_{v1}=0$
but nonzero $g_{v2}, g_{b1}$ and $g_{b2}$ as well as specification of the borderline
dimension $d_c$.  A physical region of the RG fixed point lies
below the $a\epsilon=2$ line, see in figure~4. This point completely losses stability
below the critical value of dimension $d_c=2.36$ (independently on the $a$-parameter)
and also below the value of $a_c=0.146$ (independently on the dimension).
Thus we confirm, in particular, that thermal fluctuations of the magnetic scaling
regime may occur, and, in comparison with earlier results our value of the borderline
dimension ($d_c=2.36$) is significantly lower than in the $\epsilon$ expansion
\cite{FoSuPo82} ($d_c=2.85$) and rather lower than in the 'modified' double expansion
introduced in \cite{HnHoJu01} ($d_c=2.46$) but it is rather higher then value
($d_c=2.2$) calculated in the frame of the McComb's renormalization \cite{Verma04}.

Note that the stability of any regime determines the concrete Alfven ratio $r_A$
(ratio of kinetic and magnetic energy density in MHD turbulence, see
\cite{Verma01a,Verma01b}, for example).
Once the stationary scaling regime becomes and stands, the Alfven ratio is fixed
(i.e., it means that the fixed point is reached in the field RG terminology).
Thus the injected energy necessary to steady the stationary scaling regime must
have specific value, or, in another words, all "coupling constants" $g_i$ are
fixed in scaling regime with values which are dependent on dimension $d$.
In like manner the inverse Prandtl number $u\equiv\eta/\nu$ ($\eta$ is magnetic
resistivity) is thus fixed.
Verma \cite{Verma01b} has obtained $\eta/\nu=0.85/0.36=2.36$ in $3$-dimensions
for large $r_A\approx5000$ (corresponding to region of the kinetic regime) and
for zeroth normalized cross-helicity. For smaller $r_A$ this ratio decrease to
$0.69$ for $r_A=1$, and, the both $\eta$ and $\nu$ vary approximately as $d^{-1/2}$
\cite{Verma01a}. We have mentioned above that in our double expansion calculation
in the kinetic point we have fixed the ratio $u\equiv\eta/\nu$ with its $d$-dependence
showed in figure 1.
The magnetic fixed point is characterized by decreasing value of $u$ to zero
what is in correspondence with results of \cite{Verma01b}: his calculation gives
for decreasing $r_A$ (magnetic regime) in $3$-dimensions also decreasing value
of $\eta/\nu$ as one can expect in the magnetic fixed point.

\ack %\section*{Acknowledgement}
This work was supported by Science and Technology Assistance
Agency under the contract No. APVT-51-027904, and by grant
RFFI-RFBR 05-02-17603, and by SAS, project No. 2/6193.

% ======================== BIBLIOGRAPHY ===========================

\end{document}